\documentclass{aa}    

\usepackage{graphicx,url,twoopt,natbib}
\usepackage[varg]{txfonts}           
\usepackage{hyperref}                
\usepackage{xcolor}

\hypersetup{
  colorlinks=true,   
  urlcolor=blue,     
  linkcolor=blue     
}

\bibpunct{(}{)}{;}{a}{}{,}    
\makeatletter
\newcommand{\bibnote}[2]{\global\@namedef{#1note}{#2}}
\newcommand{\biblink}[2]{\global\@namedef{#1link}{#2}}
\makeatother

\makeatletter
  \protected\def\stonyslink{%
     \def\hyper@linkstart##1##2{}\let\hyper@linkend\@empty}
  \newcommandtwoopt{\citeads}[3][][]{%
   \href{http://ui.adsabs.harvard.edu/abs/#3/abstract}%
        {\stonyslink \citealp[#1][#2]{#3}}
   \biblink{#3}{\href{http://ui.adsabs.harvard.edu/abs/#3/abstract}{ADS}}}
 \newcommandtwoopt{\citepads}[3][][]{%
   \href{http://ui.adsabs.harvard.edu/abs/#3/abstract}%
        {\stonyslink \citep[#1][#2]{#3}}
   \biblink{#3}{\href{http://ui.adsabs.harvard.edu/abs/#3/abstract}{ADS}}}
 \newcommandtwoopt{\citetads}[3][][]{%
   \href{http://ui.adsabs.harvard.edu/abs/#3/abstract}%
        {\stonyslink \citet[#1][#2]{#3}}
  \biblink{#3}{\href{http://ui.adsabs.harvard.edu/abs/#3/abstract}{ADS}}}
 \newcommandtwoopt{\citeyearads}[3][][]{%
   \href{http://ui.adsabs.harvard.edu/abs/#3/abstract}%
        {\stonyslink \citeyear[#1][#2]{#3}}
   \biblink{#3}{\href{http://ui.adsabs.harvard.edu/abs/#3/abstract}{ADS}}}
\makeatother

\newcommand{\thedate}{\today}
\definecolor{mygray}{gray}{0.6}

\begin{document}  


\twocolumn[{%
\vspace*{4ex}
\begin{center}
  {\Large \bf Secular aberration drift in stellar proper motions: An additional term due to the change in line-of-sight direction}\\[4ex] 
  {\large \bf Niu Liu$^{1}$, Zi Zhu$^{1,2}$, and Jia-Cheng Liu$^{1}$
      }\\
  {\tiny Last updated on \thedate} \\ [1em]
  \begin{minipage}[t]{16cm} \small
        $^1$ 
        School of Astronomy and Space Science,
        Key Laboratory of Modern Astronomy and Astrophysics (Ministry of Education),
        Nanjing University, Nanjing, P. R. China\\
        $^{\enspace}$ 
        e-mail: \href{mailto:niu.liu@nju.edu.cn}{\tt niu.liu@nju.edu.cn} \\
        $^2$
        University of Chinese Academy of Sciences, Nanjing 211135, China\\[4ex]
  {\bf Abstract.~} 
     The motion of the Solar System barycenter (SSB), the spatial origin of the International Celestial Reference System, causes a directional displacement known as secular aberration.
   The secular aberration drift caused by the galactocentric acceleration of the SSB has been modeled in the third generation of the International Celestial Reference Frame.   
   We aim to address another secular aberration drift effect due to the change in the line-of-sight direction and study its implications for stellar proper motions.
   We derived a complete formula for the secular aberration drift and computed its influence on stellar proper motion based on the astrometric data in \textit{Gaia} Data Release 3. 
   We find that the secular aberration drift due to the change in the line-of-sight direction tends to decrease the observed proper motions for stars with galactic longitudes between $0^{\circ}$ and $180^{\circ}$, and increase the observed proper motion for stars in the remaining region.
   If this secular aberration drift effect is ignored, it will induce an additional proper motion of $>1\,\mathrm{mas\,yr^{-1}}$ for 84 stars and $>0.02\,\mathrm{mas\,yr^{-1}}$ for 5\,944\,879 stars, which is comparable to or several times greater than the typical formal uncertainty of the \textit{Gaia} proper motion measurements at $G<13$. 
   The secular aberration drift due to the change in the line-of-sight direction and the acceleration of the SSB should be modeled to make the stellar reference frame consistent with the extragalactic reference frame.
   \vspace*{2ex}
  \end{minipage}
\end{center}
}]


\section{Introduction}

The International Celestial Reference System \citepads[ICRS;][]{1998A&A...331L..33F} adopts the Solar System barycenter (SSB) as the spatial origin and uses extragalactic sources as the fiducial points in the sky.
It has long been established that the motion of the SSB in space causes a displacement in the source direction, known as the secular aberration effect \citepads[see, e.g., the review in][]{2021A&A...649A...9G}.
This effect was previously assumed to be constant and thus was not considered in astrometric data reduction.
This assumption relies on two underlying premises:
(i) the SSB moves uniformly in space, and
(ii) the direction of the line of sight for a given source does not change.
However, these two premises do not always hold valid.

First, Galactic rotation suggests that the SSB does not exhibit uniform motion.
\citetads{2003A&A...404..743K} first formulated the secular aberration drift due to the acceleration of the SSB; according to their calculations, this drift causes an apparent proper motion field of extragalactic sources with an amplitude of approximately 5 microarcsecond per year ($\mathrm{\mu as\ yr^{-1}}$).
This effect, also referred to as Galactic aberration \citepads[e.g.][]{2023RNAAS...7..133M}, has been detected in the astrometric measurements of extragalactic sources by very long baseline interferometry \citepads[VLBI;][]{2019A&A...630A..93M} and \textit{Gaia} \citepads{2021A&A...649A...9G}.
The Galactic aberration effect results in an apparent rotation of the ICRS \citepads{2012A&A...548A..50L} and then a non-negligible drift of the celestial pole \citepads{2022A&A...665A.121Y}.
In the construction of the third generation of the International Celestial Reference Frame \citepads[ICRF3;][]{2020A&A...644A.159C}, the VLBI delay model integrates the effect of the galactocentric acceleration of the SSB.
This new model corrects for the observational effect due to the curvature of the SSB motion, making the ICRF3 more consistent with the ICRS concept than its predecessors.

Second, the assumption that the direction of the line of sight remains constant does not always hold. 
Due to their enormous distances, extragalactic sources are not supposed to show any proper motion with respect to the current observational accuracy and precision.
This is also one of the basic assumptions behind the ICRS concept.
The Galactic objects, however, show non-negligible proper motions, which alter the direction of the line of sight and, consequently, the magnitude of the secular aberration.
This effect manifests as additional proper motion and is, as we will demonstrate, proportional to the stellar proper motion.
Therefore, this effect is several orders of magnitude greater than that caused by the acceleration of the SSB and will bias the proper motion measurement for sources with high proper motion.

This Letter aims to address the secular aberration drift in stellar proper motions due to both the acceleration of the SSB and the change in the direction of the line of sight. 
The complete formula of the secular aberration drift effect is developed in Sect.~\ref{sect:secular-aberration}.
In Sect.~\ref{sect:result} we determine the implications for stellar proper motions based on \textit{Gaia} Data Release 3 \citepads[DR3;][]{2016A&A...595A...1G,2023A&A...674A...1G}.

\section{Secular aberration drift effect} \label{sect:secular-aberration}

For a fictitious observer at the SSB who is at rest with respect to the extragalactic background, the coordinate direction of an object at the observational epoch, ${T,}$ is denoted as the unit vector, $\vec{u}$.
The difference in the coordinate direction caused by the motion of the SSB can be expanded as\footnote{In this paper, we use $\vec{A}^{\prime}\vec{B}$ to denote the dot product between two vectors, $\vec{A}$ and $\vec{B}$.}
\begin{equation} \label{eq:secular-aberration}
        \delta \vec{u} = \dfrac{1}{c} \left[ \vec{V}_{\odot} - \left(\vec{u}^{\prime}\vec{V}_{\odot}\right)\vec{u} \right] + O~(c^{-2}),
\end{equation}
where $\vec{V}_{\odot}$ is the velocity vector of the SSB with respect to the extragalactic background at ${T}$, and $c$ is the speed of light.
This formula ignores the relativistic aberrational effect, that is, terms higher than the second order with respect to $1/c$.

Following the standard model of stellar motion \citepads{1997ESASP1200.....E,2016A&A...595A...4L}, the position and motion of the source can be described by six astrometric parameters.
Five of these parameters are the classical parameters: the right ascension ($\alpha$), declination ($\delta$), trigonometric parallax ($\varpi$), and the proper motion in the right ascension ($\mu_{\alpha^*}$) and declination ($\mu_{\delta}$). 
For the sixth parameter, we used the radial proper motion ($\mu_r$) instead of the radial velocity ($v_r$), which is defined as
\begin{equation}
    \mu_r = \dfrac{v_r \varpi}{A},
\end{equation}
where $A$ is the astronomical unit expressed in $\mathrm{km\,s^{-1}\,yr}$.
The normal triad at the location of this source can then be defined as
\begin{equation} \label{eq:triad}
    \vec{p} = \left( \begin{array}{c} 
        -\sin\alpha\\
        \cos\alpha \\
        0
   \end{array} \right),
    \vec{q} = \left( \begin{array}{c} 
        -\sin\delta\cos\alpha \\
        -\sin\delta\sin\alpha \\
        \cos\delta
    \end{array} \right),
    \vec{r} = \left( \begin{array}{c} 
        \cos\delta\cos\alpha \\
        \cos\delta\sin\alpha \\
        \sin\delta
    \end{array} \right).
\end{equation}
Therefore, the proper motion of the source can be expressed as 
\begin{equation}
    \vec{\mu} = \vec{p} \mu_{\alpha^*} + \vec{q} \mu_{\delta}.
\end{equation}

Since the six parameters change with time, these parameters are conventionally given with respect to a reference epoch, $T_{\rm ref}$.
We subsequently used the difference between the reference epoch and the observation epoch, $t={T}_{\rm obs}-T_{\rm ref}$, as the time argument and a subscript 0 to denote the quantities at the reference epoch (i.e., $t\!=\!0$).
Let us introduce a factor, $f,$ to account for the change in the distance as
\begin{equation}
    f = \left[ 1 + 2\mu_{r 0} t + \left( \mu^2_{0}  + \mu^2_{r 0} \right) t^2 \right]^{-1/2}
\end{equation}
and a sum vector of the (tangential) proper motion and the radial proper motion as
\begin{equation}
    \vec{\mu}_{v 0} = \vec{r}_0 \mu_{r 0} + \vec{\mu}_0 = \vec{r}_0 \mu_{r 0} + \vec{p}_0 \mu_{\alpha^* 0} + \vec{q}_0 \mu_{\delta 0}.
\end{equation}
The propagated coordinate direction, $\vec{u,}$ at $t$ can thus be written as
\begin{equation} \label{eq:stellar-motion}
    \vec{u} =  \left( \vec{r}_0 \ + \vec{\mu}_{v 0} t \right)  f.
\end{equation}
The light-time effect is ignored since it can cause only small displacements of $\mathrm{0.1\,mas}$ over 100\,yr \citepads{2014A&A...570A..62B} and hence is negligible for the investigation in this work.
The Galactic rotation of the observed stars, as in previous studies \citepads[e.g.][]{2013MNRAS.433.3597L}, is also ignored here. 

Considering the acceleration term (e.g., the Galactic rotation), the velocity of the SSB can be written as 
\begin{equation} \label{eq:ssb-velocity}
    \vec{V}_{\odot} = \vec{V}_0 + \vec{a} t,
\end{equation}
where $\vec{V}_0$ is the initial velocity vector of the SSB at $T_{\rm ref}$.
We adopted the results of the Galactic kinematic analysis in \citetads{2019ApJ...885..131R}; other studies yielded consistent results \citepads[e.g.,][]{2021MNRAS.502.4377B}.
The velocity of the SSB is
\begin{equation}
    \vec{V}_{0} =
\mathbf{A}_{G} \left(\begin{array}{c}
         U_{\odot}  \\
         -\Theta_{\odot}-V_{\odot} \\
        W_{\odot}  
     \end{array} \right) = \mathbf{A}_{G} \left(
     \begin{array}{c}
          10.6  \\
          -246.7  \\
          7.6 
     \end{array} \right) \,\mathrm{km\,s^{-1}},
\end{equation}
where $\mathbf{A}_{G}$ is the transform matrix between the equatorial and galactic systems \citepads[see][Vol. 1, Sect. 1.5.3]{1997ESASP1200.....E}.
The distance to the Galactic center, $R_0$, is approximately $\mathrm{8.15~kpc}$.
The acceleration of the SSB is dominated by the galactocentric acceleration.

By substituting Eqs.~(\ref{eq:stellar-motion}) and (\ref{eq:ssb-velocity}) into Eq.~(\ref{eq:secular-aberration}), the secular aberration effect can be expanded as
\begin{equation} \label{eq:secular-aberration-expansion}
\begin{aligned}
\delta \vec{u}  
&= \dfrac{1}{c} \left[ \vec{V}_{0} - \left( \vec{r}_0^{\prime} \vec{V}_0  \right) \vec{r}_0 f^2 \right] \\
&+ \dfrac{t}{c} \left[ \vec{a} - \left( \vec{r}_0^{\prime} \vec{a} \right) \vec{r}_0 f^2 \right] - \dfrac{f^2 t}{c} \left[ \left( \vec{\mu}_{v 0}^{\prime} \vec{V}_0 \right) \vec{r}_0 + \left( \vec{r}_0^{\prime} \vec{V}_0 \right) \vec{\mu}_{v 0} \right] \\
&-\dfrac{f^2 t^2}{c} \left[ \left( \vec{r}_0^{\prime}\vec{a} + \vec{\mu}_{v 0}^{\prime}\vec{V}_0 \right) \vec{\mu}_{v 0} + \left( \vec{\mu}^{\prime}_{v 0} \vec{a} \right) \vec{r}_0 \right] 
- \dfrac{f^2 t^3}{c} \left( \vec{\mu}^{\prime}_{v 0} \vec{a} \right) \vec{\mu}_{v 0}.
\end{aligned}
\end{equation}
The secular aberration effect apparently changes with time due to both the acceleration of the SSB and the change in the line-of-sight direction of the source.

For extragalactic sources, the proper motion and radial proper motion are nominally assumed to be zero, which leads to $f\!=\!1$.
As a result, Eq.~(\ref{eq:secular-aberration-expansion}) becomes
\begin{equation} \label{eq:sad-quasar}
    \delta \vec{u} = \dfrac{1}{c} \left[ \vec{V}_{0} - \left( \vec{r}_0^{\prime} \vec{V}_0  \right) \vec{r}_0 \right] + \dfrac{t}{c} \left[ \vec{a} - \left( \vec{r}_0^{\prime} \vec{a} \right) \vec{r}_0 \right].
\end{equation}
The first term in Eq.~(\ref{eq:sad-quasar}) represents the constant secular aberration, while the second term is caused by the acceleration of the SSB and manifests as a dipole field of the apparent proper motion.
Using the values given in \citetads{2019ApJ...885..131R} in the computation, we have
\begin{equation} \label{eq:constant-secular-aberration}
 \dfrac{V_0}{c} = 8.2 \times 10^{-4} = 169.855^{\prime\prime}
\end{equation}
for constant aberration displacement and
\begin{equation} \label{eq:ga-quasar}
        \dfrac{a}{c} \simeq \dfrac{V_0^2}{R_0 c} = \mathrm{5.3\,\mu as\ yr^{-1}}\\
\end{equation}
for the apparent proper motion.
Although the magnitude of the constant aberration greatly exceeds the accuracy of microarcsecond astrometry, the effect itself does not vary with time; thus, it is omitted in practical data analyses.
The effect of the acceleration of the SSB depends on the location of the source, resulting in a dipolar proper motion field with an amplitude of approximately $\mathrm{5\,\mu as\ yr^{-1}}$.
This effect has been well studied and will not be discussed further in this work.

As the proper motions of the Galactic stars are non-negligible, the variation in the secular aberration effect is more complicated.
Here, we were more concerned with the secular aberration drift since it will be absorbed in the observed stellar proper motion if it is neglected.
For a source with an extremely high proper motion of $\mathrm{10\,arcsec\,yr^{-1}}$, the amplitudes of the quadratic and cubic terms of $t$ could reach $\mathrm{0.04~\mu as\,yr^{-2}}$ and $\mathrm{1.2\times10^{-8}~\mu as\,yr^{-3}}$, respectively.
These two terms might need to be considered for astrometric data reductions that cover a long time baseline; they can be safely omitted in this work.
We used the Taylor expansions up to the first order to approximate $f^2$, which yields
\begin{equation} \label{eq:f2-expansion}
    f^2 = 1 - 2\mu_{r 0} t - \left( \mu^2_0 + \mu^2_{r 0} \right) t^2 + \cdots.
\end{equation}
Then, we substituted Eq.~(\ref{eq:f2-expansion}) into Eq.~(\ref{eq:secular-aberration-expansion}) and kept only the constant and linear terms of $t$, leading to
\begin{equation} \label{eq:sad_star}
\begin{aligned}
    \delta \vec{u} &= \dfrac{1}{c} \left[ \vec{V}_{0} - \left( \vec{r}_0^{\prime} \vec{V}_0  \right) \vec{r}_0 \right] + \dfrac{t}{c} \left[ \vec{a} - \left( \vec{r}_0^{\prime} \vec{a} \right) \vec{r}_0 \right] \\
    &-\dfrac{t}{c} \left[ \left( \vec{\mu}_{v0}^{\prime} \vec{V}_0 \right) \vec{r}_0 + \left( \vec{r}_0^{\prime} \vec{V}_0 \right) \vec{\mu}_{v0} \right] + \cdots
\end{aligned}
\end{equation}

This equation suggests that the apparent proper motion due to the secular aberration effect contains an additional term compared to extragalactic sources.
This term can be regarded as an effect related to the change in the line-of-sight direction of the source, whose magnitude not only depends on the celestial coordinate but is also proportional to the proper motion of the source by a factor of $V_0 / c \simeq 8.0 \times 10^{-4}$.
In \textit{Gaia} DR3, the source with the largest total proper motion is Barnard's star (\textit{Gaia} DR3 4472832130942575872), whose total proper motion is approximately $\mathrm{10393.349\,mas\,yr^{-1}}$ with a formal uncertainty of $\mathrm{0.04\,mas\,yr^{-1}}$.
For this star, the magnitude of the third term in Eq.~(\ref{eq:sad_star}) reaches $\mathrm{8\,mas\,yr^{-1}}$, which is three orders of magnitude greater than the magnitude of the second term in Eq.~(\ref{eq:sad_star}) and two orders of magnitude greater than the formal uncertainty of the \textit{Gaia} proper motion measurement.
This kind of secular aberration drift effect, therefore, should be taken into consideration. 
Conversely, with a required accuracy of $\mathrm{1~\mu as\,yr^{-1}}$ and $\mathrm{10~\mu as\,yr^{-1}}$ in proper motion, the secular aberration drift due to the proper motion of a source may need to be considered if the stellar proper motion reaches $\mathrm{1.2\,mas\,yr^{-1}}$ and $\mathrm{12\,mas\,yr^{-1}}$, respectively.

In short, the secular aberration drift induces two additional terms in the observed proper motion for the stars, one caused by the acceleration of the SSB and the other caused by the proper motion of the source.
%
The first term has been well investigated and determined in previous studies; however, the second term is rarely discussed in the literature.
In this work, we focus on the second term.

\begin{figure}[!htpb]
  \centering
  \includegraphics[trim={0 2.5cm 2.5cm 2.5cm},clip,width=\columnwidth]{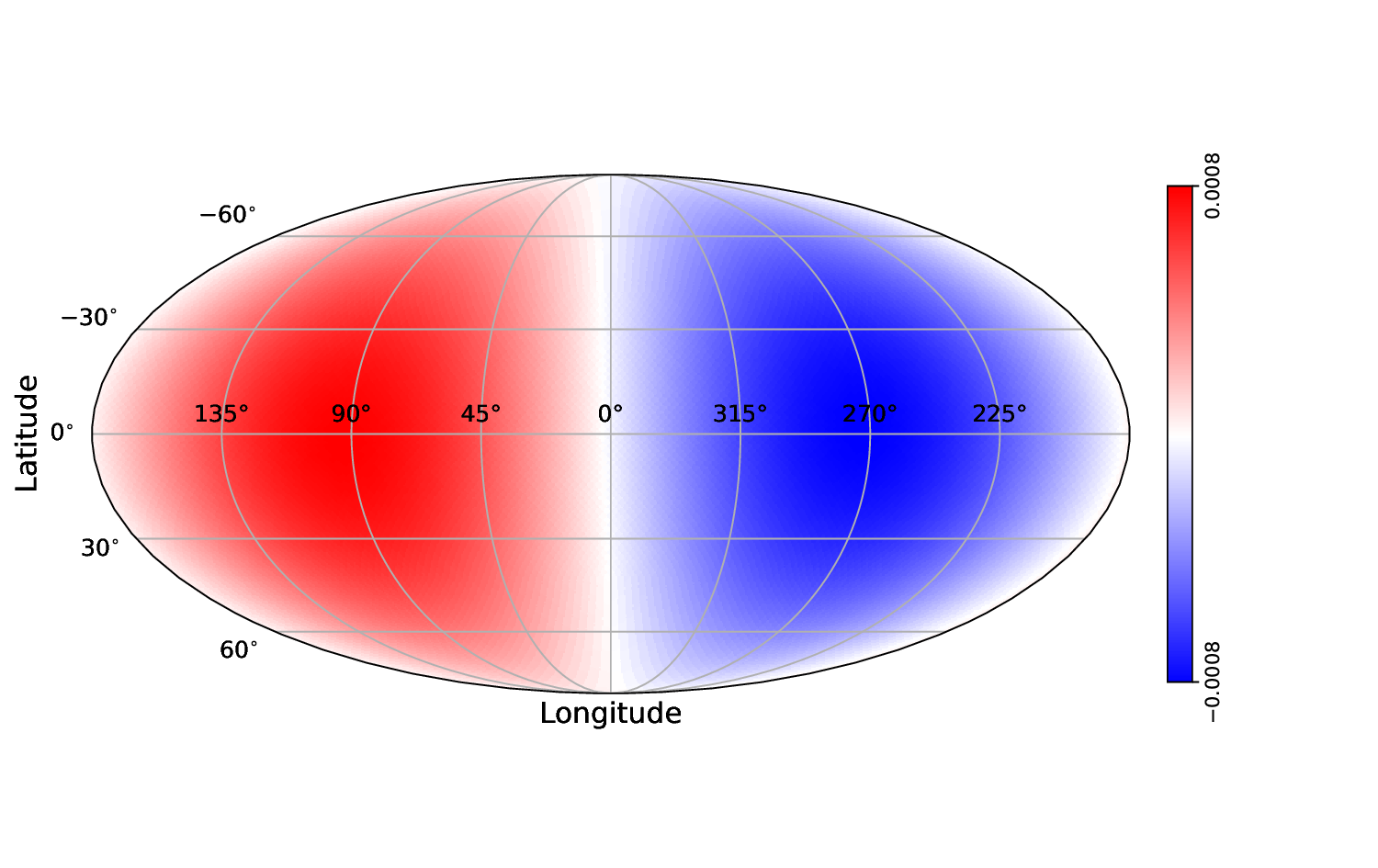}
  \caption{\label{fig:pm_fac}
  Distribution of the factor $-(\vec{r}_0^{\prime} \vec{V}_0 ) / c$ in the galactic coordinate system, representing the relative change in the observed total proper motion due to secular aberration drift caused by the stellar proper motion.
  Blue indicates regions where the total proper motion will decrease, and red indicates regions where it will increase.}
\end{figure}

We note that the terms proportional to $\vec{r}_0$ in Eq.~(\ref{eq:sad_star}) do not lead to proper motion changes.
Therefore, the additional proper motion caused by the secular aberration drift due to the change in the line-of-sight direction can be written as
\begin{equation} \label{eq:sad_star_pm_vec}
    \Delta \vec{\mu} = -\dfrac{1}{c} \left( \vec{r}_0^{\prime} \vec{V}_0 \right) \vec{\mu}_{0}.
\end{equation}
This equation suggests that this effect will only change the magnitude of the proper motion by a factor of $1-( \vec{r}_0^{\prime} \vec{V}_0 ) / c$.
The corresponding components in right ascension and declination are
%
    \begin{align}
       \Delta \mu_{\alpha^*} &= -\dfrac{1}{c}  \left( \vec{r}_0^{\prime} \vec{V}_0 \right)  \mu_{\alpha^* 0}, \label{eq:sad_pm_ra} \\
       \Delta \mu_{\delta} &= -\dfrac{1}{c} \left( \vec{r}_0^{\prime} \vec{V}_0 \right)  \mu_{\delta 0}. \label{eq:sad_pm_dec} 
    \end{align}
%
Figure~\ref{fig:pm_fac} displays the relative change in the observed total proper motion, according to Eq.~(\ref{eq:sad_star_pm_vec}), in the galactic coordinate system.
This pattern suggests that secular aberration drift will cause the observed proper motion to decrease for stars of $0<l<180^{\circ}$ but increase in the region of $180^{\circ}<l<360^{\circ}$.

\section{Implication for stellar proper motion} \label{sect:result}

\begin{figure}[!htpb]
  \centering
  \includegraphics[width=\columnwidth]{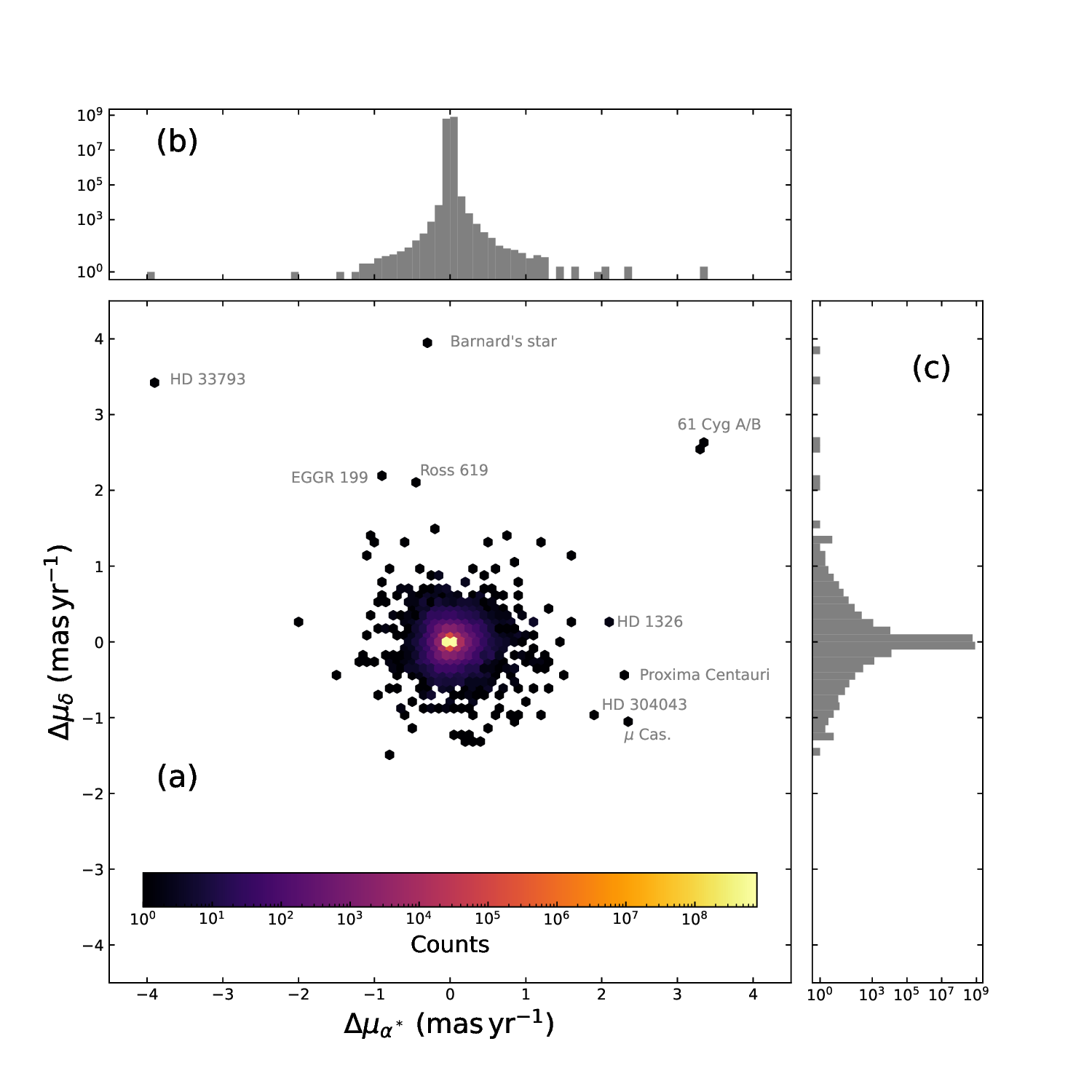}
  \caption{\label{fig:pmra_pmdec_scatter}
  Distribution of the additional proper motion components in right ascension ($\Delta \mu_{\alpha^*}$) and declination ($\Delta \mu_\delta$) caused by secular aberration drift due to the change in the line of sight for Galactic stars with a total proper motion exceeding $\mathrm{1\,mas\,yr^{-1}}$.
  (a) Scatter plot of $\Delta \mu_{\alpha^*}$ versus $\Delta \mu_\delta$. The ten sources with the greatest additional proper motion are highlighted, with their names labeled near the data points.
  (b) Histogram of $\Delta \mu_{\alpha^*}$. 
  (c) Histogram of $\Delta \mu_{\delta}$. 
   }
\end{figure}

\begin{figure}[!htpb]
  \centering
  \includegraphics[width=\columnwidth]{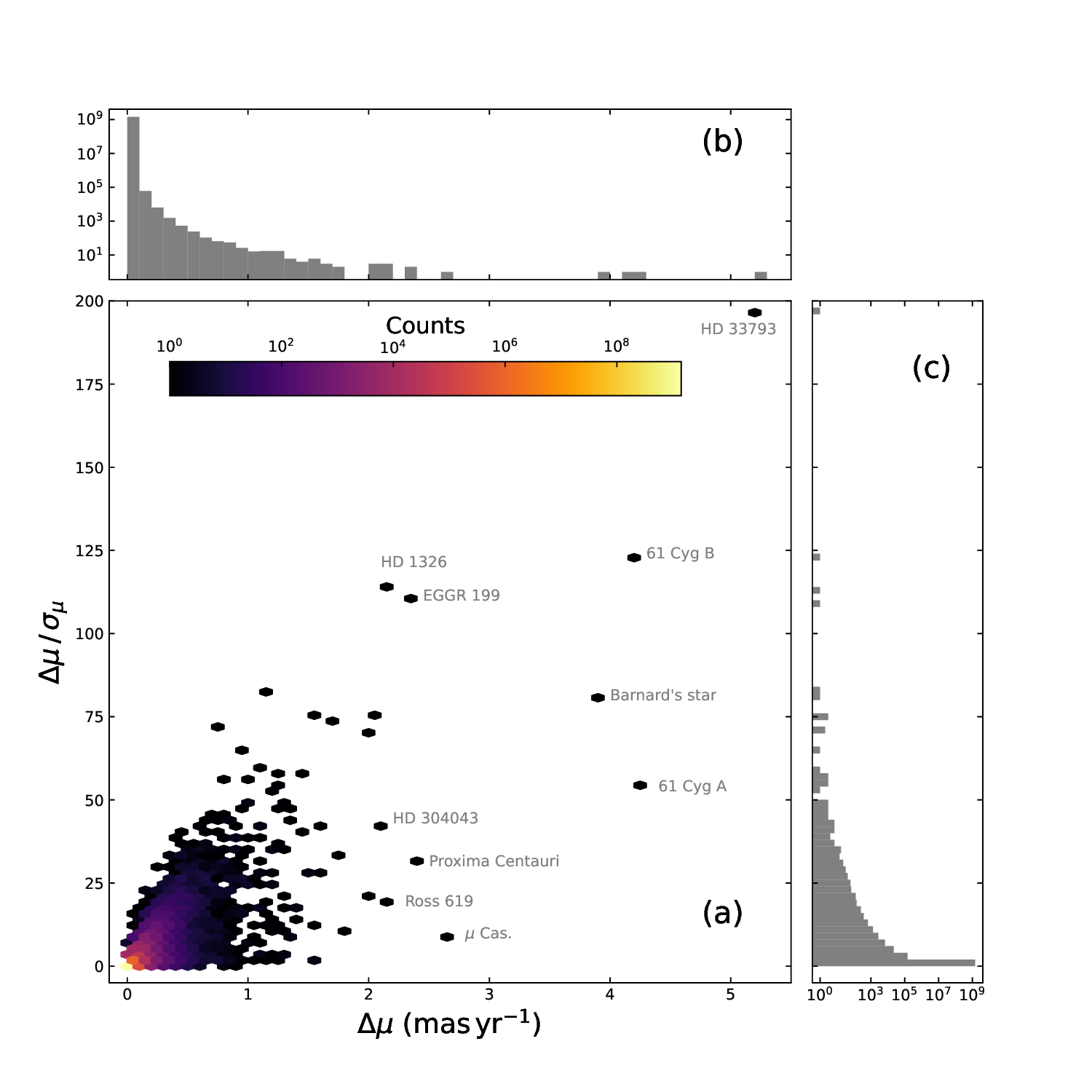}
  \caption{\label{fig:pm_pmsig_scatter}
  Distribution of the additional total proper motions ($\Delta \mu$) and their significance ($\Delta \mu / \sigma_\mu$) caused by secular aberration drift due to the change in the line of sight for Galactic stars with a total proper motion exceeding $\mathrm{1\,mas\,yr^{-1}}$.
  (a) Scatter plot of $\Delta \mu$ versus $\Delta \mu / \sigma_\mu$. The ten sources with the greatest additional proper motion are highlighted, with their names labeled near the data points.
  (b) Histogram of $\Delta \mu$. 
  (c) Histogram of $\Delta \mu / \sigma_\mu$. 
   }
\end{figure}

\begin{figure}[!htpb]
  \centering
  \includegraphics[width=\columnwidth]{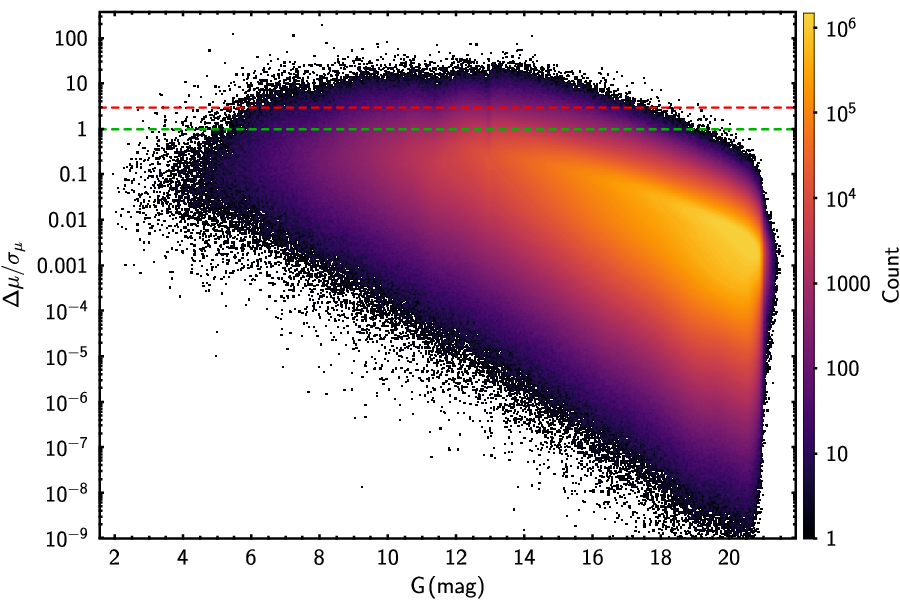}
  \caption{\label{fig:pmsig_vs_G}
  Distribution of the significance of the additional total proper motion ($\Delta \mu / \sigma_\mu$) caused by secular aberration drift due to the change in the line of sight as a function of \textit{Gaia} $G$ magnitude for Galactic stars with a total proper motion exceeding $\mathrm{1\,mas\,yr^{-1}}$.
  The horizontal lines in green and red are at vertical axis values of 1 and 3, respectively. 
   }
\end{figure}

\begin{figure}[!htpb]
  \centering
  \includegraphics[width=\columnwidth]{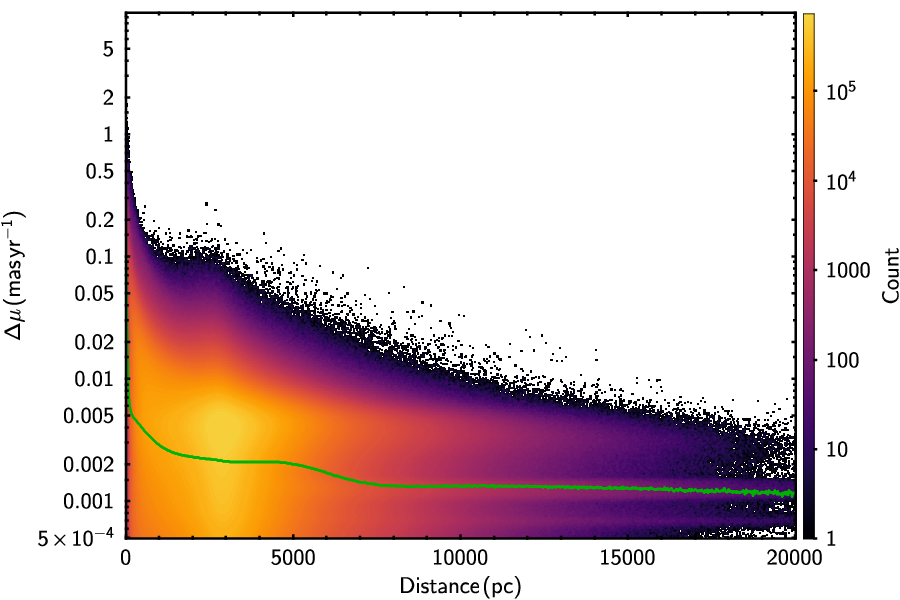}
  \caption{\label{fig:pm_d_scatter}
   Distribution of the additional total proper motions caused by secular aberration drift due to the change in the line of sight as a function of distances for Galactic stars with total proper motions exceeding $\mathrm{1\,mas\,yr^{-1}}$.
   The green line indicates the medians, calculated using the built-in Epanechnikov kernel smoothing function in the TOPCAT package \citepads{2011ascl.soft01010T}.}
\end{figure}

We searched for stellar objects (i.e., with $\texttt{in\_qso\_candidates}\!=\!\mathrm{false}$ and $\texttt{in\_galaxy\_candidates}\!=\!\mathrm{false}$) that had either a five- or a six-parameter solution and a total proper motion, $\mu_{0}$, greater than $\mathrm{1\,mas\,yr^{-1}}$ ($\texttt{pm} > 1$) in  \textit{Gaia} DR3.
The query returned a sample of 1\,417\,503\,136 sources.
We then computed the additional proper motion due to secular aberration drift using Eqs.~(\ref{eq:sad_star_pm_vec})--(\ref{eq:sad_pm_dec}).
Figure~\ref{fig:pmra_pmdec_scatter} depicts the distribution of additional proper motion components.
The greatest additional proper motion is $\mathrm{5.214\,mas\,yr^{-1}}$ for \textit{Gaia} DR3 4810594479418041856 (HD 33793).
The number of stars whose additional proper motion exceeds $\mathrm{1\,mas\,yr^{-1}}$ in an absolute sense is 43 in right ascension, 29 in declination, and 84 for the total proper motion.
Considering that the typical formal uncertainty of the \textit{Gaia} proper motion at $G < 13$ is approximately $\mathrm{0.02\,mas\,yr^{-1}}$ \citepads{2021A&A...649A...2L}, the corresponding numbers of stars with additional proper motion exceeding $\mathrm{0.02\,mas\,yr^{-1}}$ are  2\,875\,610 for right ascension, 2\,422\,948 for declination, and 5\,944\,879 for total proper motion.
The uncertainties of the proper motions in the next \textit{Gaia} data release will be improved by at least  a factor of 0.35 compared to those in \textit{Gaia} DR3 \citepads{2021A&A...649A...2L}, making the additional proper motion effect more pronounced.

To quantify the significance of the additional proper motion, we computed the ratio between the additional proper motions and the formal uncertainties of \textit{Gaia} proper motion measurements.
The formal uncertainty of the \textit{Gaia} total proper motion was computed as
\begin{equation}
    \sigma_\mu = \sqrt{\sigma^2_{\mu\alpha^*} + \sigma^2_{\mu\delta} + 2 C_{\mu_{\alpha^*},\mu_{\delta}} \sigma_{\mu\alpha^*} \sigma_{\mu\delta}},
\end{equation}
where $\sigma_{\mu\alpha^*}$ and $\sigma_{\mu\delta}$ are the uncertainties in proper motion in right ascension and declination, respectively, and $C_{\mu_{\alpha^*},\mu_{\delta}}$ is the covariance between them.

As shown in Fig.~\ref{fig:pm_pmsig_scatter}, large values of the additional total proper motion are usually associated with large values of significance.
For more than 70\,000 stars, the additional total proper motion is significant at the 3$\sigma$ level or higher and thus cannot be ignored for these stars.
Figure~\ref{fig:pmsig_vs_G} depicts the distribution of the significance of the additional total proper motion as a function of \textit{Gaia} G magnitude, clearly showing that significant additional proper motions due to the secular aberration drift mainly occur in the magnitude range $6<G<16$.

We examined the dependence of the additional proper motion on distance.
Using the built-in function for distance estimation in the STILTS package \citepads{2006ASPC..351..666T}, we computed stellar distances from \textit{Gaia} parallaxes.
This estimation is based on the exponentially decreasing space density prior defined in \citetads{2016ApJ...833..119A}; we set the length scale, $L,$ to 1.35\,kpc as done therein.
Figure~\ref{fig:pm_d_scatter} depicts the dependence of the additional proper motion as a function of the stellar distance.
We find that the additional proper motion generally decreases with distance.
Moreover, for stars within 2\,kpc, the additional proper motion can reach up to $\mathrm{0.1\,mas\,yr^{-1}}$, which highlights the potential impact of secular aberration drift on Galactic kinematic analyses, particularly for nearby stars.

\section{Discussion}

In the construction of the ICRF3, the secular aberration effect due to the galactocentric acceleration of the SSB has been modeled, enhancing the consistency of ICRF3 with the concept of the ICRS.
It is anticipated that the \textit{Gaia} celestial reference frame will also model and correct for this effect.
To ensure that the stellar reference frame of \textit{Gaia} aligns with the extragalactic reference frame, it is necessary to account for the secular aberration drift due to stellar proper motion.

For the extragalactic frame, it is sufficient to consider the galactocentric acceleration of the SSB. 
However, the stellar frame must address the secular aberration drift caused by stellar proper motion.
As demonstrated in the previous section, the secular aberration drift from stellar proper motion can be several times greater than the effect of the SSB's acceleration and the formal uncertainties in \textit{Gaia} proper motion measurements, particularly for bright stars with high proper motions.
Therefore, we recommend that the term ``Galactic aberration effect,'' commonly used in the ICRF community, encompass the secular aberration drift in (apparent) proper motions due to both the SSB's acceleration and stellar proper motion to ensure consistency between the stellar frame and the extragalactic frame.

Fully correcting the stellar proper motion terms related to secular aberration drift requires knowledge of the position and motion of the SSB within our Galaxy.
However, this information is usually derived from Galactic kinematic analyses, which utilize stellar proper motion data.
As a result, correcting for secular aberration drift in stellar proper motion necessitates an iterative process. 
We propose addressing this issue in four steps:\\
(i) Determine the acceleration vector of the SSB from the proper motion measurements of extragalactic sources, and then correct for the secular aberration drift due to the acceleration of the SSB for both the Galactic and extragalactic sources.\\
(ii) Adopt the state-of-the-art estimates from the literature as the initial guess for the velocity of the SSB and correct for the secular aberration drift due to the proper motion for Galactic stars.\\
(iii) Perform a Galactic kinematic analysis using the corrected stellar proper motions to determine the velocity of the SSB; \\
(iv) Repeat steps (ii) and (iii) until the results converge to the required accuracy.\\
By following these steps, we can achieve a more accurate and consistent celestial reference frame that accounts for both the SSB’s acceleration and the proper motions of stars.

\section{Conclusions}

We investigated the secular aberration drift due to stellar proper motion in addition to the effect caused by the acceleration of the Solar System.
Our findings demonstrate that this previously neglected term can induce systematic errors in observed stellar proper motion, depending on both the celestial position and the magnitude of the stellar proper motion.
For several dozen stars with high proper motion, this drift can bias the \textit{Gaia} proper motion by $1\,\mathrm{mas\,yr^{-1}}$ or more.
Such a bias exceeds the precision of \textit{Gaia}'s proper motion measurements at $G<13$ and should therefore be considered in the reduction of \textit{Gaia} astrometric data, especially for stars at $6<G<16$.
 
By fully accounting for the secular aberration drift in proper motion, we can bring the stellar and extragalactic reference frames more in line with the definition of the ICRS.
This correction is crucial for maintaining a consistency between the stellar frame and the extragalactic frame and for advancing our understanding of stellar kinematics.

\begin{acknowledgements} 
    We sincerely thank the anonymous referee for their constructive comments and useful suggestions, which improve the work a lot.
    N. Liu, J.-C. Liu and Z. Zhu were supported by the National Natural Science Foundation of China (NSFC) under grant Nos~12373074, 11833004, and 12103026.
    This work has made use of data from the European Space Agency (ESA) mission
{\it Gaia} (\url{https://www.cosmos.esa.int/gaia}), processed by the {\it Gaia}
Data Processing and Analysis Consortium (DPAC,
\url{https://www.cosmos.esa.int/web/gaia/dpac/consortium}). Funding for the DPAC
has been provided by national institutions, in particular the institutions
participating in the {\it Gaia} Multilateral Agreement.
    This research had also made use of several Python packages, including Numpy \citepads{2011CSE....13b..22V}, IPython \citepads{2007CSE.....9c..21P}, Astropy\footnote{\url{http://www.astropy.org}} \citepads{2013A&A...558A..33A,2018AJ....156..123A}, the 2D plotting library Matplotlib \citepads{2007CSE.....9...90H}, healpy, and HEALPix\footnote{\url{http://healpix.sf.net}}.
    All the necessary files (e.g., codes) to reproduce the results in the this paper can be found at \href{https://git.nju.edu.cn/astrometry/sad_on_pm}{https://git.nju.edu.cn/astrometry/sad\_on\_pm}.
\end{acknowledgements}

\bibliographystyle{aa-note} 
\bibliography{main}      

\begin{thebibliography}{25}
\expandafter\ifx\csname natexlab\endcsname\relax\def\natexlab#1{#1}\fi

\bibitem[{{Astraatmadja} \& {Bailer-Jones}(2016)}]{2016ApJ...833..119A}
{Astraatmadja}, T.~L. \& {Bailer-Jones}, C. A.~L. 2016, \apj, 833, 119

\bibitem[{{Astropy Collaboration} {et~al.}(2018){Astropy Collaboration}, {Price-Whelan}, {Sip{\H{o}}cz}, {G{\"u}nther}, {Lim}, {Crawford}, {Conseil}, {Shupe}, {Craig}, {Dencheva}, {Ginsburg}, {VanderPlas}, {Bradley}, {P{\'e}rez-Su{\'a}rez}, {de Val-Borro}, {Aldcroft}, {Cruz}, {Robitaille}, {Tollerud}, {Ardelean}, {Babej}, {Bach}, {Bachetti}, {Bakanov}, {Bamford}, {Barentsen}, {Barmby}, {Baumbach}, {Berry}, {Biscani}, {Boquien}, {Bostroem}, {Bouma}, {Brammer}, {Bray}, {Breytenbach}, {Buddelmeijer}, {Burke}, {Calderone}, {Cano Rodr{\'\i}guez}, {Cara}, {Cardoso}, {Cheedella}, {Copin}, {Corrales}, {Crichton}, {D'Avella}, {Deil}, {Depagne}, {Dietrich}, {Donath}, {Droettboom}, {Earl}, {Erben}, {Fabbro}, {Ferreira}, {Finethy}, {Fox}, {Garrison}, {Gibbons}, {Goldstein}, {Gommers}, {Greco}, {Greenfield}, {Groener}, {Grollier}, {Hagen}, {Hirst}, {Homeier}, {Horton}, {Hosseinzadeh}, {Hu}, {Hunkeler}, {Ivezi{\'c}}, {Jain}, {Jenness}, {Kanarek}, {Kendrew}, {Kern}, {Kerzendorf}, {Khvalko}, {King}, {Kirkby}, {Kulkarni}, {Kumar}, {Lee}, {Lenz}, {Littlefair}, {Ma}, {Macleod}, {Mastropietro}, {McCully}, {Montagnac}, {Morris}, {Mueller}, {Mumford}, {Muna}, {Murphy}, {Nelson}, {Nguyen}, {Ninan}, {N{\"o}the}, {Ogaz}, {Oh}, {Parejko}, {Parley}, {Pascual}, {Patil}, {Patil}, {Plunkett}, {Prochaska}, {Rastogi}, {Reddy Janga}, {Sabater}, {Sakurikar}, {Seifert}, {Sherbert}, {Sherwood-Taylor}, {Shih}, {Sick}, {Silbiger}, {Singanamalla}, {Singer}, {Sladen}, {Sooley}, {Sornarajah}, {Streicher}, {Teuben}, {Thomas}, {Tremblay}, {Turner}, {Terr{\'o}n}, {van Kerkwijk}, {de la Vega}, {Watkins}, {Weaver}, {Whitmore}, {Woillez}, {Zabalza}, \& {Astropy Contributors}}]{2018AJ....156..123A}
{Astropy Collaboration}, {Price-Whelan}, A.~M., {Sip{\H{o}}cz}, B.~M., {et~al.} 2018, \aj, 156, 123

\bibitem[{{Astropy Collaboration} {et~al.}(2013){Astropy Collaboration}, {Robitaille}, {Tollerud}, {Greenfield}, {Droettboom}, {Bray}, {Aldcroft}, {Davis}, {Ginsburg}, {Price-Whelan}, {Kerzendorf}, {Conley}, {Crighton}, {Barbary}, {Muna}, {Ferguson}, {Grollier}, {Parikh}, {Nair}, {Unther}, {Deil}, {Woillez}, {Conseil}, {Kramer}, {Turner}, {Singer}, {Fox}, {Weaver}, {Zabalza}, {Edwards}, {Azalee Bostroem}, {Burke}, {Casey}, {Crawford}, {Dencheva}, {Ely}, {Jenness}, {Labrie}, {Lim}, {Pierfederici}, {Pontzen}, {Ptak}, {Refsdal}, {Servillat}, \& {Streicher}}]{2013A&A...558A..33A}
{Astropy Collaboration}, {Robitaille}, T.~P., {Tollerud}, E.~J., {et~al.} 2013, \aap, 558, A33

\bibitem[{{Bobylev} {et~al.}(2021){Bobylev}, {Bajkova}, {Rastorguev}, \& {Zabolotskikh}}]{2021MNRAS.502.4377B}
{Bobylev}, V.~V., {Bajkova}, A.~T., {Rastorguev}, A.~S., \& {Zabolotskikh}, M.~V. 2021, \mnras, 502, 4377

\bibitem[{{Butkevich} \& {Lindegren}(2014)}]{2014A&A...570A..62B}
{Butkevich}, A.~G. \& {Lindegren}, L. 2014, \aap, 570, A62

\bibitem[{{Charlot} {et~al.}(2020){Charlot}, {Jacobs}, {Gordon}, {Lambert}, {de Witt}, {B{\"o}hm}, {Fey}, {Heinkelmann}, {Skurikhina}, {Titov}, {Arias}, {Bolotin}, {Bourda}, {Ma}, {Malkin}, {Nothnagel}, {Mayer}, {MacMillan}, {Nilsson}, \& {Gaume}}]{2020A&A...644A.159C}
{Charlot}, P., {Jacobs}, C.~S., {Gordon}, D., {et~al.} 2020, \aap, 644, A159

\bibitem[{ESA(1997)}]{1997ESASP1200.....E}
ESA, ed. 1997, ESA Special Publication, Vol. 1200, {The HIPPARCOS and TYCHO catalogues. Astrometric and photometric star catalogues derived from the ESA HIPPARCOS Space Astrometry Mission}

\bibitem[{{Feissel} \& {Mignard}(1998)}]{1998A&A...331L..33F}
{Feissel}, M. \& {Mignard}, F. 1998, \aap, 331, L33

\bibitem[{{Gaia Collaboration} {et~al.}(2021){Gaia Collaboration}, {Klioner}, {Mignard}, {Lindegren}, {Bastian}, {McMillan}, {Hern{\'a}ndez}, {Hobbs}, {Ramos-Lerate}, {Biermann}, {Bombrun}, {de Torres}, {Gerlach}, {Geyer}, {Hilger}, {Lammers}, {Steidelm{\"u}ller}, {Stephenson}, {Brown}, {Vallenari}, {Prusti}, {de Bruijne}, {Babusiaux}, {Creevey}, {Evans}, {Eyer}, {Hutton}, {Jansen}, {Jordi}, {Luri}, {Panem}, {Pourbaix}, {Randich}, {Sartoretti}, {Soubiran}, {Walton}, {Arenou}, {Bailer-Jones}, {Cropper}, {Drimmel}, {Katz}, {Lattanzi}, {van Leeuwen}, {Bakker}, {Casta{\~n}eda}, {De Angeli}, {Ducourant}, {Fabricius}, {Fouesneau}, {Fr{\'e}mat}, {Guerra}, {Guerrier}, {Guiraud}, {Jean-Antoine Piccolo}, {Masana}, {Messineo}, {Mowlavi}, {Nicolas}, {Nienartowicz}, {Pailler}, {Panuzzo}, {Riclet}, {Roux}, {Seabroke}, {Sordo}, {Tanga}, {Th{\'e}venin}, {Gracia-Abril}, {Portell}, {Teyssier}, {Altmann}, {Andrae}, {Bellas-Velidis}, {Benson}, {Berthier}, {Blomme}, {Brugaletta}, {Burgess}, {Busso}, {Carry}, {Cellino}, {Cheek}, {Clementini}, {Damerdji}, {Davidson}, {Delchambre}, {Dell'Oro}, {Fern{\'a}ndez-Hern{\'a}ndez}, {Galluccio}, {Garc{\'\i}a-Lario}, {Garcia-Reinaldos}, {Gonz{\'a}lez-N{\'u}{\~n}ez}, {Gosset}, {Haigron}, {Halbwachs}, {Hambly}, {Harrison}, {Hatzidimitriou}, {Heiter}, {Hestroffer}, {Hodgkin}, {Holl}, {Jan{\ss}en}, {Jevardat de Fombelle}, {Jordan}, {Krone-Martins}, {Lanzafame}, {L{\"o}ffler}, {Lorca}, {Manteiga}, {Marchal}, {Marrese}, {Moitinho}, {Mora}, {Muinonen}, {Osborne}, {Pancino}, {Pauwels}, {Recio-Blanco}, {Richards}, {Riello}, {Rimoldini}, {Robin}, {Roegiers}, {Rybizki}, {Sarro}, {Siopis}, {Smith}, {Sozzetti}, {Ulla}, {Utrilla}, {van Leeuwen}, {van Reeven}, {Abbas}, {Abreu Aramburu}, {Accart}, {Aerts}, {Aguado}, {Ajaj}, {Altavilla}, {{\'A}lvarez}, {{\'A}lvarez Cid-Fuentes}, {Alves}, {Anderson}, {Anglada Varela}, {Antoja}, {Audard}, {Baines}, {Baker}, {Balaguer-N{\'u}{\~n}ez}, {Balbinot}, {Balog}, {Barache}, {Barbato}, {Barros}, {Barstow}, {Bartolom{\'e}}, {Bassilana}, {Bauchet}, {Baudesson-Stella}, {Becciani}, {Bellazzini}, {Bernet}, {Bertone}, {Bianchi}, {Blanco-Cuaresma}, {Boch}, {Bossini}, {Bouquillon}, {Bramante}, {Breedt}, {Bressan}, {Brouillet}, {Bucciarelli}, {Burlacu}, {Busonero}, {Butkevich}, {Buzzi}, {Caffau}, {Cancelliere}, {C{\'a}novas}, {Cantat-Gaudin}, {Carballo}, {Carlucci}, {Carnerero}, {Carrasco}, {Casamiquela}, {Castellani}, {Castro-Ginard}, {Castro Sampol}, {Chaoul}, {Charlot}, {Chemin}, {Chiavassa}, {Comoretto}, {Cooper}, {Cornez}, {Cowell}, {Crifo}, {Crosta}, {Crowley}, {Dafonte}, {Dapergolas}, {David}, {David}, {de Laverny}, {De Luise}, {De March}, {De Ridder}, {de Souza}, {de Teodoro}, {del Peloso}, {del Pozo}, {Delgado}, {Delgado}, {Delisle}, {Di Matteo}, {Diakite}, {Diener}, {Distefano}, {Dolding}, {Eappachen}, {Enke}, {Esquej}, {Fabre}, {Fabrizio}, {Faigler}, {Fedorets}, {Fernique}, {Fienga}, {Figueras}, {Fouron}, {Fragkoudi}, {Fraile}, {Franke}, {Gai}, {Garabato}, {Garcia-Gutierrez}, {Garc{\'\i}a-Torres}, {Garofalo}, {Gavras}, {Giacobbe}, {Gilmore}, {Girona}, {Giuffrida}, {Gomez}, {Gonzalez-Santamaria}, {Gonz{\'a}lez-Vidal}, {Granvik}, {Guti{\'e}rrez-S{\'a}nchez}, {Guy}, {Hauser}, {Haywood}, {Helmi}, {Hidalgo}, {H{\l}adczuk}, {Holland}, {Huckle}, {Jasniewicz}, {Jonker}, {Juaristi Campillo}, {Julbe}, {Karbevska}, {Kervella}, {Khanna}, {Kochoska}, {Kordopatis}, {Korn}, {Kostrzewa-Rutkowska}, {Kruszy{\'n}ska}, {Lambert}, {Lanza}, {Lasne}, {Le Campion}, {Le Fustec}, {Lebreton}, {Lebzelter}, {Leccia}, {Leclerc}, {Lecoeur-Taibi}, {Liao}, {Licata}, {Lindstr{\o}m}, {Lister}, {Livanou}, {Lobel}, {Madrero Pardo}, {Managau}, {Mann}, {Marchant}, {Marconi}, {Marcos Santos}, {Marinoni}, {Marocco}, {Marshall}, {Martin Polo}, {Mart{\'\i}n-Fleitas}, {Masip}, {Massari}, {Mastrobuono-Battisti}, {Mazeh}, {Messina}, {Michalik}, {Millar}, {Mints}, {Molina}, {Molinaro}, {Moln{\'a}r}, {Montegriffo}, {Mor}, {Morbidelli}, {Morel}, {Morris}, {Mulone}, {Munoz}, {Muraveva}, {Murphy}, {Musella}, {Noval}, {Ord{\'e}novic}, {Orr{\`u}}, {Osinde}, {Pagani}, {Pagano}, {Palaversa}, {Palicio}, {Panahi}, {Pawlak}, {Pe{\~n}alosa Esteller}, {Penttil{\"a}}, {Piersimoni}, {Pineau}, {Plachy}, {Plum}, {Poggio}, {Poretti}, {Poujoulet}, {Pr{\v{s}}a}, {Pulone}, {Racero}, {Ragaini}, {Rainer}, {Raiteri}, {Rambaux}, {Ramos}, {Re Fiorentin}, {Regibo}, {Reyl{\'e}}, {Ripepi}, {Riva}, {Rixon}, {Robichon}, {Robin}, {Roelens}, {Rohrbasser}, {Romero-G{\'o}mez}, {Rowell}, {Royer}, {Rybicki}, {Sadowski}, {Sagrist{\`a} Sell{\'e}s}, {Sahlmann}, {Salgado}, {Salguero}, {Samaras}, {Sanchez Gimenez}, {Sanna}, {Santove{\~n}a}, {Sarasso}, {Schultheis}, {Sciacca}, {Segol}, {Segovia}, {S{\'e}gransan}, {Semeux}, {Siddiqui}, {Siebert}, {Siltala}, {Slezak}, {Smart}, {Solano}, {Solitro}, {Souami}, {Souchay}, {Spagna}, {Spoto}, {Steele}, {S{\"u}veges}, {Szabados}, {Szegedi-Elek}, {Taris}, {Tauran}, {Taylor}, {Teixeira}, {Thuillot}, {Tonello}, {Torra}, {Torra}, {Turon}, {Unger}, {Vaillant}, {van Dillen}, {Vanel}, {Vecchiato}, {Viala}, {Vicente}, {Voutsinas}, {Weiler}, {Wevers}, {Wyrzykowski}, {Yoldas}, {Yvard}, {Zhao}, {Zorec}, {Zucker}, {Zurbach}, \& {Zwitter}}]{2021A&A...649A...9G}
{Gaia Collaboration}, {Klioner}, S.~A., {Mignard}, F., {et~al.} 2021, \aap, 649, A9

\bibitem[{{Gaia Collaboration} {et~al.}(2016){Gaia Collaboration}, {Prusti}, {de Bruijne}, {Brown}, {Vallenari}, {Babusiaux}, {Bailer-Jones}, {Bastian}, {Biermann}, {Evans}, {Eyer}, {Jansen}, {Jordi}, {Klioner}, {Lammers}, {Lindegren}, {Luri}, {Mignard}, {Milligan}, {Panem}, {Poinsignon}, {Pourbaix}, {Randich}, {Sarri}, {Sartoretti}, {Siddiqui}, {Soubiran}, {Valette}, {van Leeuwen}, {Walton}, {Aerts}, {Arenou}, {Cropper}, {Drimmel}, {H{\o}g}, {Katz}, {Lattanzi}, {O'Mullane}, {Grebel}, {Holland}, {Huc}, {Passot}, {Bramante}, {Cacciari}, {Casta{\~n}eda}, {Chaoul}, {Cheek}, {De Angeli}, {Fabricius}, {Guerra}, {Hern{\'a}ndez}, {Jean-Antoine-Piccolo}, {Masana}, {Messineo}, {Mowlavi}, {Nienartowicz}, {Ord{\'o}{\~n}ez-Blanco}, {Panuzzo}, {Portell}, {Richards}, {Riello}, {Seabroke}, {Tanga}, {Th{\'e}venin}, {Torra}, {Els}, {Gracia-Abril}, {Comoretto}, {Garcia-Reinaldos}, {Lock}, {Mercier}, {Altmann}, {Andrae}, {Astraatmadja}, {Bellas-Velidis}, {Benson}, {Berthier}, {Blomme}, {Busso}, {Carry}, {Cellino}, {Clementini}, {Cowell}, {Creevey}, {Cuypers}, {Davidson}, {De Ridder}, {de Torres}, {Delchambre}, {Dell'Oro}, {Ducourant}, {Fr{\'e}mat}, {Garc{\'\i}a-Torres}, {Gosset}, {Halbwachs}, {Hambly}, {Harrison}, {Hauser}, {Hestroffer}, {Hodgkin}, {Huckle}, {Hutton}, {Jasniewicz}, {Jordan}, {Kontizas}, {Korn}, {Lanzafame}, {Manteiga}, {Moitinho}, {Muinonen}, {Osinde}, {Pancino}, {Pauwels}, {Petit}, {Recio-Blanco}, {Robin}, {Sarro}, {Siopis}, {Smith}, {Smith}, {Sozzetti}, {Thuillot}, {van Reeven}, {Viala}, {Abbas}, {Abreu Aramburu}, {Accart}, {Aguado}, {Allan}, {Allasia}, {Altavilla}, {{\'A}lvarez}, {Alves}, {Anderson}, {Andrei}, {Anglada Varela}, {Antiche}, {Antoja}, {Ant{\'o}n}, {Arcay}, {Atzei}, {Ayache}, {Bach}, {Baker}, {Balaguer-N{\'u}{\~n}ez}, {Barache}, {Barata}, {Barbier}, {Barblan}, {Baroni}, {Barrado y Navascu{\'e}s}, {Barros}, {Barstow}, {Becciani}, {Bellazzini}, {Bellei}, {Bello Garc{\'\i}a}, {Belokurov}, {Bendjoya}, {Berihuete}, {Bianchi}, {Bienaym{\'e}}, {Billebaud}, {Blagorodnova}, {Blanco-Cuaresma}, {Boch}, {Bombrun}, {Borrachero}, {Bouquillon}, {Bourda}, {Bouy}, {Bragaglia}, {Breddels}, {Brouillet}, {Br{\"u}semeister}, {Bucciarelli}, {Budnik}, {Burgess}, {Burgon}, {Burlacu}, {Busonero}, {Buzzi}, {Caffau}, {Cambras}, {Campbell}, {Cancelliere}, {Cantat-Gaudin}, {Carlucci}, {Carrasco}, {Castellani}, {Charlot}, {Charnas}, {Charvet}, {Chassat}, {Chiavassa}, {Clotet}, {Cocozza}, {Collins}, {Collins}, {Costigan}, {Crifo}, {Cross}, {Crosta}, {Crowley}, {Dafonte}, {Damerdji}, {Dapergolas}, {David}, {David}, {De Cat}, {de Felice}, {de Laverny}, {De Luise}, {De March}, {de Martino}, {de Souza}, {Debosscher}, {del Pozo}, {Delbo}, {Delgado}, {Delgado}, {di Marco}, {Di Matteo}, {Diakite}, {Distefano}, {Dolding}, {Dos Anjos}, {Drazinos}, {Dur{\'a}n}, {Dzigan}, {Ecale}, {Edvardsson}, {Enke}, {Erdmann}, {Escolar}, {Espina}, {Evans}, {Eynard Bontemps}, {Fabre}, {Fabrizio}, {Faigler}, {Falc{\~a}o}, {Farr{\`a}s Casas}, {Faye}, {Federici}, {Fedorets}, {Fern{\'a}ndez-Hern{\'a}ndez}, {Fernique}, {Fienga}, {Figueras}, {Filippi}, {Findeisen}, {Fonti}, {Fouesneau}, {Fraile}, {Fraser}, {Fuchs}, {Furnell}, {Gai}, {Galleti}, {Galluccio}, {Garabato}, {Garc{\'\i}a-Sedano}, {Gar{\'e}}, {Garofalo}, {Garralda}, {Gavras}, {Gerssen}, {Geyer}, {Gilmore}, {Girona}, {Giuffrida}, {Gomes}, {Gonz{\'a}lez-Marcos}, {Gonz{\'a}lez-N{\'u}{\~n}ez}, {Gonz{\'a}lez-Vidal}, {Granvik}, {Guerrier}, {Guillout}, {Guiraud}, {G{\'u}rpide}, {Guti{\'e}rrez-S{\'a}nchez}, {Guy}, {Haigron}, {Hatzidimitriou}, {Haywood}, {Heiter}, {Helmi}, {Hobbs}, {Hofmann}, {Holl}, {Holland}, {Hunt}, {Hypki}, {Icardi}, {Irwin}, {Jevardat de Fombelle}, {Jofr{\'e}}, {Jonker}, {Jorissen}, {Julbe}, {Karampelas}, {Kochoska}, {Kohley}, {Kolenberg}, {Kontizas}, {Koposov}, {Kordopatis}, {Koubsky}, {Kowalczyk}, {Krone-Martins}, {Kudryashova}, {Kull}, {Bachchan}, {Lacoste-Seris}, {Lanza}, {Lavigne}, {Le Poncin-Lafitte}, {Lebreton}, {Lebzelter}, {Leccia}, {Leclerc}, {Lecoeur-Taibi}, {Lemaitre}, {Lenhardt}, {Leroux}, {Liao}, {Licata}, {Lindstr{\o}m}, {Lister}, {Livanou}, {Lobel}, {L{\"o}ffler}, {L{\'o}pez}, {Lopez-Lozano}, {Lorenz}, {Loureiro}, {MacDonald}, {Magalh{\~a}es Fernandes}, {Managau}, {Mann}, {Mantelet}, {Marchal}, {Marchant}, {Marconi}, {Marie}, {Marinoni}, {Marrese}, {Marschalk{\'o}}, {Marshall}, {Mart{\'\i}n-Fleitas}, {Martino}, {Mary}, {Matijevi{\v{c}}}, {Mazeh}, {McMillan}, {Messina}, {Mestre}, {Michalik}, {Millar}, {Miranda}, {Molina}, {Molinaro}, {Molinaro}, {Moln{\'a}r}, {Moniez}, {Montegriffo}, {Monteiro}, {Mor}, {Mora}, {Morbidelli}, {Morel}, {Morgenthaler}, {Morley}, {Morris}, {Mulone}, {Muraveva}, {Musella}, {Narbonne}, {Nelemans}, {Nicastro}, {Noval}, {Ord{\'e}novic}, {Ordieres-Mer{\'e}}, {Osborne}, {Pagani}, {Pagano}, {Pailler}, {Palacin}, {Palaversa}, {Parsons}, {Paulsen}, {Pecoraro}, {Pedrosa}, {Pentik{\"a}inen}, {Pereira}, {Pichon}, {Piersimoni}, {Pineau}, {Plachy}, {Plum}, {Poujoulet}, {Pr{\v{s}}a}, {Pulone}, {Ragaini}, {Rago}, {Rambaux}, {Ramos-Lerate}, {Ranalli}, {Rauw}, {Read}, {Regibo}, {Renk}, {Reyl{\'e}}, {Ribeiro}, {Rimoldini}, {Ripepi}, {Riva}, {Rixon}, {Roelens}, {Romero-G{\'o}mez}, {Rowell}, {Royer}, {Rudolph}, {Ruiz-Dern}, {Sadowski}, {Sagrist{\`a} Sell{\'e}s}, {Sahlmann}, {Salgado}, {Salguero}, {Sarasso}, {Savietto}, {Schnorhk}, {Schultheis}, {Sciacca}, {Segol}, {Segovia}, {Segransan}, {Serpell}, {Shih}, {Smareglia}, {Smart}, {Smith}, {Solano}, {Solitro}, {Sordo}, {Soria Nieto}, {Souchay}, {Spagna}, {Spoto}, {Stampa}, {Steele}, {Steidelm{\"u}ller}, {Stephenson}, {Stoev}, {Suess}, {S{\"u}veges}, {Surdej}, {Szabados}, {Szegedi-Elek}, {Tapiador}, {Taris}, {Tauran}, {Taylor}, {Teixeira}, {Terrett}, {Tingley}, {Trager}, {Turon}, {Ulla}, {Utrilla}, {Valentini}, {van Elteren}, {Van Hemelryck}, {van Leeuwen}, {Varadi}, {Vecchiato}, {Veljanoski}, {Via}, {Vicente}, {Vogt}, {Voss}, {Votruba}, {Voutsinas}, {Walmsley}, {Weiler}, {Weingrill}, {Werner}, {Wevers}, {Whitehead}, {Wyrzykowski}, {Yoldas}, {{\v{Z}}erjal}, {Zucker}, {Zurbach}, {Zwitter}, {Alecu}, {Allen}, {Allende Prieto}, {Amorim}, {Anglada-Escud{\'e}}, {Arsenijevic}, {Azaz}, {Balm}, {Beck}, {Bernstein}, {Bigot}, {Bijaoui}, {Blasco}, {Bonfigli}, {Bono}, {Boudreault}, {Bressan}, {Brown}, {Brunet}, {Bunclark}, {Buonanno}, {Butkevich}, {Carret}, {Carrion}, {Chemin}, {Ch{\'e}reau}, {Corcione}, {Darmigny}, {de Boer}, {de Teodoro}, {de Zeeuw}, {Delle Luche}, {Domingues}, {Dubath}, {Fodor}, {Fr{\'e}zouls}, {Fries}, {Fustes}, {Fyfe}, {Gallardo}, {Gallegos}, {Gardiol}, {Gebran}, {Gomboc}, {G{\'o}mez}, {Grux}, {Gueguen}, {Heyrovsky}, {Hoar}, {Iannicola}, {Isasi Parache}, {Janotto}, {Joliet}, {Jonckheere}, {Keil}, {Kim}, {Klagyivik}, {Klar}, {Knude}, {Kochukhov}, {Kolka}, {Kos}, {Kutka}, {Lainey}, {LeBouquin}, {Liu}, {Loreggia}, {Makarov}, {Marseille}, {Martayan}, {Martinez-Rubi}, {Massart}, {Meynadier}, {Mignot}, {Munari}, {Nguyen}, {Nordlander}, {Ocvirk}, {O'Flaherty}, {Olias Sanz}, {Ortiz}, {Osorio}, {Oszkiewicz}, {Ouzounis}, {Palmer}, {Park}, {Pasquato}, {Peltzer}, {Peralta}, {P{\'e}turaud}, {Pieniluoma}, {Pigozzi}, {Poels}, {Prat}, {Prod'homme}, {Raison}, {Rebordao}, {Risquez}, {Rocca-Volmerange}, {Rosen}, {Ruiz-Fuertes}, {Russo}, {Sembay}, {Serraller Vizcaino}, {Short}, {Siebert}, {Silva}, {Sinachopoulos}, {Slezak}, {Soffel}, {Sosnowska}, {Strai{\v{z}}ys}, {ter Linden}, {Terrell}, {Theil}, {Tiede}, {Troisi}, {Tsalmantza}, {Tur}, {Vaccari}, {Vachier}, {Valles}, {Van Hamme}, {Veltz}, {Virtanen}, {Wallut}, {Wichmann}, {Wilkinson}, {Ziaeepour}, \& {Zschocke}}]{2016A&A...595A...1G}
{Gaia Collaboration}, {Prusti}, T., {de Bruijne}, J.~H.~J., {et~al.} 2016, \aap, 595, A1

\bibitem[{{Gaia Collaboration} {et~al.}(2023){Gaia Collaboration}, {Vallenari}, {Brown}, {Prusti}, {de Bruijne}, {Arenou}, {Babusiaux}, {Biermann}, {Creevey}, {Ducourant}, {Evans}, {Eyer}, {Guerra}, {Hutton}, {Jordi}, {Klioner}, {Lammers}, {Lindegren}, {Luri}, {Mignard}, {Panem}, {Pourbaix}, {Randich}, {Sartoretti}, {Soubiran}, {Tanga}, {Walton}, {Bailer-Jones}, {Bastian}, {Drimmel}, {Jansen}, {Katz}, {Lattanzi}, {van Leeuwen}, {Bakker}, {Cacciari}, {Casta{\~n}eda}, {De Angeli}, {Fabricius}, {Fouesneau}, {Fr{\'e}mat}, {Galluccio}, {Guerrier}, {Heiter}, {Masana}, {Messineo}, {Mowlavi}, {Nicolas}, {Nienartowicz}, {Pailler}, {Panuzzo}, {Riclet}, {Roux}, {Seabroke}, {Sordo}, {Th{\'e}venin}, {Gracia-Abril}, {Portell}, {Teyssier}, {Altmann}, {Andrae}, {Audard}, {Bellas-Velidis}, {Benson}, {Berthier}, {Blomme}, {Burgess}, {Busonero}, {Busso}, {C{\'a}novas}, {Carry}, {Cellino}, {Cheek}, {Clementini}, {Damerdji}, {Davidson}, {de Teodoro}, {Nu{\~n}ez Campos}, {Delchambre}, {Dell'Oro}, {Esquej}, {Fern{\'a}ndez-Hern{\'a}ndez}, {Fraile}, {Garabato}, {Garc{\'\i}a-Lario}, {Gosset}, {Haigron}, {Halbwachs}, {Hambly}, {Harrison}, {Hern{\'a}ndez}, {Hestroffer}, {Hodgkin}, {Holl}, {Jan{\ss}en}, {Jevardat de Fombelle}, {Jordan}, {Krone-Martins}, {Lanzafame}, {L{\"o}ffler}, {Marchal}, {Marrese}, {Moitinho}, {Muinonen}, {Osborne}, {Pancino}, {Pauwels}, {Recio-Blanco}, {Reyl{\'e}}, {Riello}, {Rimoldini}, {Roegiers}, {Rybizki}, {Sarro}, {Siopis}, {Smith}, {Sozzetti}, {Utrilla}, {van Leeuwen}, {Abbas}, {{\'A}brah{\'a}m}, {Abreu Aramburu}, {Aerts}, {Aguado}, {Ajaj}, {Aldea-Montero}, {Altavilla}, {{\'A}lvarez}, {Alves}, {Anders}, {Anderson}, {Anglada Varela}, {Antoja}, {Baines}, {Baker}, {Balaguer-N{\'u}{\~n}ez}, {Balbinot}, {Balog}, {Barache}, {Barbato}, {Barros}, {Barstow}, {Bartolom{\'e}}, {Bassilana}, {Bauchet}, {Becciani}, {Bellazzini}, {Berihuete}, {Bernet}, {Bertone}, {Bianchi}, {Binnenfeld}, {Blanco-Cuaresma}, {Blazere}, {Boch}, {Bombrun}, {Bossini}, {Bouquillon}, {Bragaglia}, {Bramante}, {Breedt}, {Bressan}, {Brouillet}, {Brugaletta}, {Bucciarelli}, {Burlacu}, {Butkevich}, {Buzzi}, {Caffau}, {Cancelliere}, {Cantat-Gaudin}, {Carballo}, {Carlucci}, {Carnerero}, {Carrasco}, {Casamiquela}, {Castellani}, {Castro-Ginard}, {Chaoul}, {Charlot}, {Chemin}, {Chiaramida}, {Chiavassa}, {Chornay}, {Comoretto}, {Contursi}, {Cooper}, {Cornez}, {Cowell}, {Crifo}, {Cropper}, {Crosta}, {Crowley}, {Dafonte}, {Dapergolas}, {David}, {David}, {de Laverny}, {De Luise}, {De March}, {De Ridder}, {de Souza}, {de Torres}, {del Peloso}, {del Pozo}, {Delbo}, {Delgado}, {Delisle}, {Demouchy}, {Dharmawardena}, {Di Matteo}, {Diakite}, {Diener}, {Distefano}, {Dolding}, {Edvardsson}, {Enke}, {Fabre}, {Fabrizio}, {Faigler}, {Fedorets}, {Fernique}, {Fienga}, {Figueras}, {Fournier}, {Fouron}, {Fragkoudi}, {Gai}, {Garcia-Gutierrez}, {Garcia-Reinaldos}, {Garc{\'\i}a-Torres}, {Garofalo}, {Gavel}, {Gavras}, {Gerlach}, {Geyer}, {Giacobbe}, {Gilmore}, {Girona}, {Giuffrida}, {Gomel}, {Gomez}, {Gonz{\'a}lez-N{\'u}{\~n}ez}, {Gonz{\'a}lez-Santamar{\'\i}a}, {Gonz{\'a}lez-Vidal}, {Granvik}, {Guillout}, {Guiraud}, {Guti{\'e}rrez-S{\'a}nchez}, {Guy}, {Hatzidimitriou}, {Hauser}, {Haywood}, {Helmer}, {Helmi}, {Sarmiento}, {Hidalgo}, {Hilger}, {H{\l}adczuk}, {Hobbs}, {Holland}, {Huckle}, {Jardine}, {Jasniewicz}, {Jean-Antoine Piccolo}, {Jim{\'e}nez-Arranz}, {Jorissen}, {Juaristi Campillo}, {Julbe}, {Karbevska}, {Kervella}, {Khanna}, {Kontizas}, {Kordopatis}, {Korn}, {K{\'o}sp{\'a}l}, {Kostrzewa-Rutkowska}, {Kruszy{\'n}ska}, {Kun}, {Laizeau}, {Lambert}, {Lanza}, {Lasne}, {Le Campion}, {Lebreton}, {Lebzelter}, {Leccia}, {Leclerc}, {Lecoeur-Taibi}, {Liao}, {Licata}, {Lindstr{\o}m}, {Lister}, {Livanou}, {Lobel}, {Lorca}, {Loup}, {Madrero Pardo}, {Magdaleno Romeo}, {Managau}, {Mann}, {Manteiga}, {Marchant}, {Marconi}, {Marcos}, {Marcos Santos}, {Mar{\'\i}n Pina}, {Marinoni}, {Marocco}, {Marshall}, {Martin Polo}, {Mart{\'\i}n-Fleitas}, {Marton}, {Mary}, {Masip}, {Massari}, {Mastrobuono-Battisti}, {Mazeh}, {McMillan}, {Messina}, {Michalik}, {Millar}, {Mints}, {Molina}, {Molinaro}, {Moln{\'a}r}, {Monari}, {Mongui{\'o}}, {Montegriffo}, {Montero}, {Mor}, {Mora}, {Morbidelli}, {Morel}, {Morris}, {Muraveva}, {Murphy}, {Musella}, {Nagy}, {Noval}, {Oca{\~n}a}, {Ogden}, {Ordenovic}, {Osinde}, {Pagani}, {Pagano}, {Palaversa}, {Palicio}, {Pallas-Quintela}, {Panahi}, {Payne-Wardenaar}, {Pe{\~n}alosa Esteller}, {Penttil{\"a}}, {Pichon}, {Piersimoni}, {Pineau}, {Plachy}, {Plum}, {Poggio}, {Pr{\v{s}}a}, {Pulone}, {Racero}, {Ragaini}, {Rainer}, {Raiteri}, {Rambaux}, {Ramos}, {Ramos-Lerate}, {Re Fiorentin}, {Regibo}, {Richards}, {Rios Diaz}, {Ripepi}, {Riva}, {Rix}, {Rixon}, {Robichon}, {Robin}, {Robin}, {Roelens}, {Rogues}, {Rohrbasser}, {Romero-G{\'o}mez}, {Rowell}, {Royer}, {Ruz Mieres}, {Rybicki}, {Sadowski}, {S{\'a}ez N{\'u}{\~n}ez}, {Sagrist{\`a} Sell{\'e}s}, {Sahlmann}, {Salguero}, {Samaras}, {Sanchez Gimenez}, {Sanna}, {Santove{\~n}a}, {Sarasso}, {Schultheis}, {Sciacca}, {Segol}, {Segovia}, {S{\'e}gransan}, {Semeux}, {Shahaf}, {Siddiqui}, {Siebert}, {Siltala}, {Silvelo}, {Slezak}, {Slezak}, {Smart}, {Snaith}, {Solano}, {Solitro}, {Souami}, {Souchay}, {Spagna}, {Spina}, {Spoto}, {Steele}, {Steidelm{\"u}ller}, {Stephenson}, {S{\"u}veges}, {Surdej}, {Szabados}, {Szegedi-Elek}, {Taris}, {Taylor}, {Teixeira}, {Tolomei}, {Tonello}, {Torra}, {Torra}, {Torralba Elipe}, {Trabucchi}, {Tsounis}, {Turon}, {Ulla}, {Unger}, {Vaillant}, {van Dillen}, {van Reeven}, {Vanel}, {Vecchiato}, {Viala}, {Vicente}, {Voutsinas}, {Weiler}, {Wevers}, {Wyrzykowski}, {Yoldas}, {Yvard}, {Zhao}, {Zorec}, {Zucker}, \& {Zwitter}}]{2023A&A...674A...1G}
{Gaia Collaboration}, {Vallenari}, A., {Brown}, A.~G.~A., {et~al.} 2023, \aap, 674, A1

\bibitem[{{Hunter}(2007)}]{2007CSE.....9...90H}
{Hunter}, J.~D. 2007, Computing in Science and Engineering, 9, 90

\bibitem[{{Kovalevsky}(2003)}]{2003A&A...404..743K}
{Kovalevsky}, J. 2003, \aap, 404, 743

\bibitem[{{Lindegren} {et~al.}(2021){Lindegren}, {Klioner}, {Hern{\'a}ndez}, {Bombrun}, {Ramos-Lerate}, {Steidelm{\"u}ller}, {Bastian}, {Biermann}, {de Torres}, {Gerlach}, {Geyer}, {Hilger}, {Hobbs}, {Lammers}, {McMillan}, {Stephenson}, {Casta{\~n}eda}, {Davidson}, {Fabricius}, {Gracia-Abril}, {Portell}, {Rowell}, {Teyssier}, {Torra}, {Bartolom{\'e}}, {Clotet}, {Garralda}, {Gonz{\'a}lez-Vidal}, {Torra}, {Abbas}, {Altmann}, {Anglada Varela}, {Balaguer-N{\'u}{\~n}ez}, {Balog}, {Barache}, {Becciani}, {Bernet}, {Bertone}, {Bianchi}, {Bouquillon}, {Brown}, {Bucciarelli}, {Busonero}, {Butkevich}, {Buzzi}, {Cancelliere}, {Carlucci}, {Charlot}, {Cioni}, {Crosta}, {Crowley}, {del Peloso}, {del Pozo}, {Drimmel}, {Esquej}, {Fienga}, {Fraile}, {Gai}, {Garcia-Reinaldos}, {Guerra}, {Hambly}, {Hauser}, {Jan{\ss}en}, {Jordan}, {Kostrzewa-Rutkowska}, {Lattanzi}, {Liao}, {Licata}, {Lister}, {L{\"o}ffler}, {Marchant}, {Masip}, {Mignard}, {Mints}, {Molina}, {Mora}, {Morbidelli}, {Murphy}, {Pagani}, {Panuzzo}, {Pe{\~n}alosa Esteller}, {Poggio}, {Re Fiorentin}, {Riva}, {Sagrist{\`a} Sell{\'e}s}, {Sanchez Gimenez}, {Sarasso}, {Sciacca}, {Siddiqui}, {Smart}, {Souami}, {Spagna}, {Steele}, {Taris}, {Utrilla}, {van Reeven}, \& {Vecchiato}}]{2021A&A...649A...2L}
{Lindegren}, L., {Klioner}, S.~A., {Hern{\'a}ndez}, J., {et~al.} 2021, \aap, 649, A2

\bibitem[{{Lindegren} {et~al.}(2016){Lindegren}, {Lammers}, {Bastian}, {Hern{\'a}ndez}, {Klioner}, {Hobbs}, {Bombrun}, {Michalik}, {Ramos-Lerate}, {Butkevich}, {Comoretto}, {Joliet}, {Holl}, {Hutton}, {Parsons}, {Steidelm{\"u}ller}, {Abbas}, {Altmann}, {Andrei}, {Anton}, {Bach}, {Barache}, {Becciani}, {Berthier}, {Bianchi}, {Biermann}, {Bouquillon}, {Bourda}, {Br{\"u}semeister}, {Bucciarelli}, {Busonero}, {Carlucci}, {Casta{\~n}eda}, {Charlot}, {Clotet}, {Crosta}, {Davidson}, {de Felice}, {Drimmel}, {Fabricius}, {Fienga}, {Figueras}, {Fraile}, {Gai}, {Garralda}, {Geyer}, {Gonz{\'a}lez-Vidal}, {Guerra}, {Hambly}, {Hauser}, {Jordan}, {Lattanzi}, {Lenhardt}, {Liao}, {L{\"o}ffler}, {McMillan}, {Mignard}, {Mora}, {Morbidelli}, {Portell}, {Riva}, {Sarasso}, {Serraller}, {Siddiqui}, {Smart}, {Spagna}, {Stampa}, {Steele}, {Taris}, {Torra}, {van Reeven}, {Vecchiato}, {Zschocke}, {de Bruijne}, {Gracia}, {Raison}, {Lister}, {Marchant}, {Messineo}, {Soffel}, {Osorio}, {de Torres}, \& {O'Mullane}}]{2016A&A...595A...4L}
{Lindegren}, L., {Lammers}, U., {Bastian}, U., {et~al.} 2016, \aap, 595, A4

\bibitem[{{Liu} {et~al.}(2012){Liu}, {Capitaine}, {Lambert}, {Malkin}, \& {Zhu}}]{2012A&A...548A..50L}
{Liu}, J.~C., {Capitaine}, N., {Lambert}, S.~B., {Malkin}, Z., \& {Zhu}, Z. 2012, \aap, 548, A50

\bibitem[{{Liu} {et~al.}(2013){Liu}, {Xie}, \& {Zhu}}]{2013MNRAS.433.3597L}
{Liu}, J.~C., {Xie}, Y., \& {Zhu}, Z. 2013, \mnras, 433, 3597

\bibitem[{{MacMillan} {et~al.}(2019){MacMillan}, {Fey}, {Gipson}, {Gordon}, {Jacobs}, {Kr{\'a}sn{\'a}}, {Lambert}, {Malkin}, {Titov}, {Wang}, \& {Xu}}]{2019A&A...630A..93M}
{MacMillan}, D.~S., {Fey}, A., {Gipson}, J.~M., {et~al.} 2019, \aap, 630, A93

\bibitem[{{Malkin}(2023)}]{2023RNAAS...7..133M}
{Malkin}, Z. 2023, Research Notes of the American Astronomical Society, 7, 133

\bibitem[{{Perez} \& {Granger}(2007)}]{2007CSE.....9c..21P}
{Perez}, F. \& {Granger}, B.~E. 2007, Computing in Science and Engineering, 9, 21

\bibitem[{{Reid} {et~al.}(2019){Reid}, {Menten}, {Brunthaler}, {Zheng}, {Dame}, {Xu}, {Li}, {Sakai}, {Wu}, {Immer}, {Zhang}, {Sanna}, {Moscadelli}, {Rygl}, {Bartkiewicz}, {Hu}, {Quiroga-Nu{\~n}ez}, \& {van Langevelde}}]{2019ApJ...885..131R}
{Reid}, M.~J., {Menten}, K.~M., {Brunthaler}, A., {et~al.} 2019, \apj, 885, 131

\bibitem[{{Taylor}(2011)}]{2011ascl.soft01010T}
{Taylor}, M. 2011, {TOPCAT: Tool for OPerations on Catalogues And Tables}, Astrophysics Source Code Library, record ascl:1101.010

\bibitem[{{Taylor}(2006)}]{2006ASPC..351..666T}
{Taylor}, M.~B. 2006, in Astronomical Society of the Pacific Conference Series, Vol. 351, Astronomical Data Analysis Software and Systems XV, ed. C.~{Gabriel}, C.~{Arviset}, D.~{Ponz}, \& S.~{Enrique}, 666

\bibitem[{{van der Walt} {et~al.}(2011){van der Walt}, {Colbert}, \& {Varoquaux}}]{2011CSE....13b..22V}
{van der Walt}, S., {Colbert}, S.~C., \& {Varoquaux}, G. 2011, Computing in Science and Engineering, 13, 22

\bibitem[{{Yao} {et~al.}(2022){Yao}, {Liu}, {Liu}, {Malkin}, {Zhu}, {Huda}, \& {Lambert}}]{2022A&A...665A.121Y}
{Yao}, J., {Liu}, J.~C., {Liu}, N., {et~al.} 2022, \aap, 665, A121

\end{thebibliography}

\end{document}